\RequirePackage{fix-cm}
\documentclass[twocolumn]{svjour3}          % twocolumn
\smartqed  % flush right qed marks, e.g. at end of proof
\usepackage{graphicx}
\usepackage{amsmath}
\usepackage{amssymb}
\usepackage{bm}
\usepackage{mathptmx}      % use Times fonts if available on your TeX system
\journalname{Nonlinear Dynamics}
\begin{document}

\title{Synchronization structures in the chain of rotating pendulums\thanks{
Theoretical results were supported by Russian Science Foundation, Project No.~19-12-00367. Numerical calculations were supported by the Ministry of Science and Higher Education of Russian Federation, Project No.~0729-2020-0036. Results presented in Appendix were supported by the Russian Foundation for Basic Research (Grant No.~18-29-10068\underline{~}mk).
%Grants or other notes
%about the article that should go on the front page should be
%placed here. General acknowledgments should be placed at the end of the article.}
}}
%\subtitle{Do you have a subtitle?\\ If so, write it here}

%\titlerunning{Short form of title}        % if too long for running head

\author{Vyacheslav~O.~Munyaev \and
		Dmitry~S.~Khorkin \and
		Maxim~I.~Bolotov \and
		Lev~A.~Smirnov \and
		Grigory~V.~Osipov %etc.
}

%\authorrunning{Short form of author list} % if too long for running head

\institute{V.~O.~Munyayev \at
              Department of Control Theory, Nizhny Novgorod State University, Gagarin~Av.~23, Nizhny~Novgorod, 603950, Russia \\
              %Tel.: +123-45-678910\\
              %Fax: +123-45-678910\\
              \email{munyaev@itmm.unn.ru}           %  \\
%             \emph{Present address:} of F. Author  %  if needed
           \and
           D.~S.~Khorkin \at
           Scientific and Educational Mathematical Center ``Mathematics of Future Technologies'', Nizhny Novgorod State University, Gagarin~Av.~23, Nizhny~Novgorod, 603950, Russia \\
           %Tel.: +123-45-678910\\
           %Fax: +123-45-678910\\
           \email{dmitryhorkin@gmail.com}
           \and
           M.~I.~Bolotov \at
           Department of Control Theory, Nizhny Novgorod State University, Gagarin~Av.~23, Nizhny~Novgorod, 603950, Russia \\
           %Tel.: +123-45-678910\\
           %Fax: +123-45-678910\\
           \email{maksim.bolotov@itmm.unn.ru}
           \and
           L.~A.~Smirnov \at
           Scientific and Educational Mathematical Center ``Mathematics of Future Technologies'', Nizhny Novgorod State University, Gagarin~Av.~23, Nizhny~Novgorod, 603950, Russia 
           \at
           Institute of Applied Physics, Russian Academy of Sciences, Ul’yanova Str.~46, Nizhny Novgorod, 603950, Russia\\
           %Tel.: +123-45-678910\\
           %Fax: +123-45-678910\\
           \email{smirnov\underline{ }lev@appl.sci-nnov.ru}
           \and
           G.~V.~Osipov \at
           Department of Control Theory, Nizhny Novgorod State University, Gagarin~Av.~23, Nizhny~Novgorod, 603950, Russia \\
           %Tel.: +123-45-678910\\
           %Fax: +123-45-678910\\
           \email{osipov@vmk.unn.ru}
}

\date{Received: date / Accepted: date}
% The correct dates will be entered by the editor

\maketitle

\begin{abstract}
The collective behavior of the ensembles of coupled nonlinear oscillator is one of the most interesting and important problems in modern nonlinear dynamics. In this paper, we study rotational dynamics, in particular space-time structures, in locally coupled identical pendulum-type elements chains that describe the behavior of phase-locked-loop systems, distributed Josephson junctions, coupled electrical machines, etc. The control parameters in the considered chains are: dissipation, coupling strength, and number of elements. In the system under consideration, the realized modes are synchronous in frequency and synchronous (in-phase) or asynchronous (out-of-phase) in phase. In the low dissipation case, the in-phase synchronous rotational regime instability region boundaries are theoretically found and the bifurcations leading to the loss of its stability are determined. The analysis was carried out for chains of arbitrary length. The existence of various out-of-phase synchronous rotational modes types is revealed: completely asynchronous in phases and the cluster in-phase synchronization regime. Regularities of transitions from one type of out-of-phase synchronous mode to another are established. It was found that at certain coupling parameter values, the coexistence of stable in-phase and out-of-phase synchronous modes is possible. It was found that for arbitrary chain length, the number of possible stable out-of-phase modes is always one less than the chain elements number. Analytical results are confirmed by numerical simulations.
\keywords{Coupled pendulums \and Synchronization \and Out-of-phase rotation \and Symmetry breaking \and Clusters}
% \PACS{PACS code1 \and PACS code2 \and more}
% \subclass{MSC code1 \and MSC code2 \and more}
\end{abstract}

\section{Introduction}
\label{intro}
The collective dynamics study in oscillatory networks of various nature is one of the actively developing topics in modern nonlinear dynamics.
\par
A general phenomenon of collective behavior is synchronization~\cite{Osipov-07,Pikovsky-01,Afraimovich-94}. Even a weak coupling strength between elements in an ensemble can lead to frequency and phase readjusting of oscillators, i.e., to synchronization.
However, this is not always the case. Due to the symmetry loss phenomenon in an identical elements population, while there is a completely synchronous regime, solutions with differing from each other elements states can be realized~\cite{Motter-10}. Systems of coupled pendulums are one of the actual models in different fields of science and techniques. Despite the relative simplicity of such models, they adequately describe not only mechanical objects but also different processes in superconducting structures~\cite{Barone-82}, molecular biology~\cite{Yakushevich-04}, and in systems of phase synchronization~\cite{Afraimovich-94}. This model is used in the study of coupled Josephson junctions dynamics~\cite{Pikovsky-01,Barone-82,Braun-04,Belykh-771,Belykh-772}. Note that the ensemble of pendulums coupled through the sine function of phase differences can be considered as the standard Kuramoto model generalization which takes into account the inertia and intrinsic nonlinearity of the population elements~\cite{Ji-14,Ha-14,Belykh-16,Belykh-20}. Among the out-of-phase regimes, the in-phase synchronization modes of pendulums in separate groups are distinguished. For example, a mode in which in-phase pendulums are located in two (three, four, etc.) parts of the chain.
\par
The aim of this paper is to study the features of rotational dynamics in chains of locally coupled identical pendulums. A detailed analytical and numerical study of the existence and stability of the in-phase rotational periodic motion and modes responsible for its instability onset has been carried out. The mechanisms of the various types out-of-phase rotational modes appearance and disappearance in chains of arbitrary length are investigated depending on the dissipation parameters and the coupling strength between the elements.
This paper is organized as follows. Sect.~\ref{sec:Model} contains a model description which is under the study and the control parameters area of our interest, in which an in-phase rotational mode exists. In Sect.~\ref{sec:InPhaseMode}, the in-phase rotational motion stability analysis is considered, the modes responsible for the instability onset are determined, as well as the corresponding asymptotic expressions for the instability regions boundaries. In Sect.~\ref{sec:OutOfPhaseModes}, out-of-phase rotational modes arising as a result of in-phase mode instability development and corresponding bifurcations are considered, their type is determined. Further, the analytical results are confirmed by the direct numerical modeling of the origin system. A summary of the main results can be found in Sect.~\ref{sec:Conclusion}.
In Appendix~\ref{sec:Eigenproblem}, an analytical description of the modes responsible for the in-phase rotational motion instability is presented. Also in Appendix~\ref{sec:Typemode} the derivation of a formula for determining the type of out-of-phase modes  is explicated.

\section{The model}
\label{sec:Model}
We consider $N$ (index $n = 1, 2, \dots, N$) coupled identical pendulums described by the system of ordinary differential equations
\begin{equation}
\ddot{\varphi}_n+\lambda\dot{\varphi}_n+\sin{\varphi_n}=\gamma+\sum_{\tilde{n}=1}^{N}K_{n\tilde{n}}\sin{\left(\varphi_{\tilde{n}}-\varphi_n\right)},
\label{eq:ModelGeneral}
\end{equation}
where $\lambda$ is the damping coefficient responsible for all the dissipative processes in the system, $\gamma$ is a constant external force identical for all $N$ pendulums, matrix $\bm{K}\in\mathbb{R}^{N\times N}$ elements $K_{n\tilde{n}}$ characterize the strength of interaction between elements.

The system \eqref{eq:ModelGeneral} exhibits non-trivial behavior for certain relationships between the parameters $\lambda$, $\gamma$ and $\bm{K}$. First of all, it should be noted that if the coordinates coincide $\varphi_{1}\left(t\right)$, $\varphi_{2}\left(t\right)$, $\ldots$, $\varphi_{N}\left(t\right)$, the system will demonstrate in-phase dynamics, i.e. $\varphi_{1}\left(t\right)=\varphi_{2}\left(t\right)=\ldots=\varphi_{N}\left(t\right)=\phi\left(t\right)$. The pendulums move in phase and their dynamics is described by the equation:
\begin{equation}
\ddot{\phi}+\lambda \dot{\phi}+\sin\phi=\gamma.
\label{eq:Pendula}
\end{equation}
The such a system behavior is well known \cite{Andronov-66}.
Further, we are interested in rotational dynamic modes, so we will dwell in detail on the parameters plane $(\lambda, \gamma)$ region, when there is periodic rotational trajectory in the phase space of the system~\eqref{eq:Pendula} and, respectively, in-phase rotary motion exists in the system~\eqref{eq:ModelGeneral}. This parameters set is approximately described by the inequalities $\gamma > T(\lambda) = 4 \lambda / \pi - 0.305 \lambda^3$ when $\lambda < \lambda^* \approx 1.22$ and $\gamma > 1$ if $\lambda > \lambda^*$ \cite{Belykh-16}, where $T(\lambda)$ defines the Tricomi bifurcation curve \cite{Tricomi-33}.

In previous works we considered the rotational dynamics features of the system~\eqref{eq:ModelGeneral} particular variations. Two coupled pendulums were investigated in \cite{Smirnov-16,Khorkin-20} for the cases of symmetric and asymmetric coupling, respectively. The three locally coupled pendulum chain was studied in \cite{Bolotov-19}, and the case of $N$ globally coupled pendulums was explored in \cite{Bolotov-20}.

Next, we will focus on the following special case of the system \eqref{eq:ModelGeneral}, namely, on the case of a chain with free ends formed by $N$ (index $n=1,2,\dots,N$) locally and identically coupled pendulums, those $K_{n\tilde{n}}=K\left(\delta_{n+1,\tilde{n}}+\delta_{n,\tilde{n}+1}\right)$, where $\delta_{n,m}$ is the Kronecker symbol ($\delta_{n,m} = 1$, if $n=m$, otherwise $\delta_{n,m} = 0$). For $N\ge 3$ the system \eqref{eq:ModelGeneral} takes the following form:
\begin{equation}
\ddot{\varphi}_n\!+\!\lambda\dot{\varphi}_n\!+\!\sin{\varphi_n}\!=\!\gamma\!+\!K\bigl(\!\sin\!{\left(\varphi_{n\!-\!1}\!-\!\varphi_n\right)}\!+\!\sin{\left(\varphi_{n\!+\!1}\!-\!\varphi_n\right)}\!\bigr)\!,
\label{eq:ModelChainLocal}
\end{equation}
where the equalities $\varphi_0=\varphi_1$, $\varphi_{N+1}=\varphi_{N}$ define the boundary conditions.

\section{In-phase rotations. Stability analysis}
\label{sec:InPhaseMode}
We linearize the system~\eqref{eq:ModelChainLocal} in the vicinity~$\phi\left(t\right)$ to determine the stability conditions for the in-phase mode, then $\varphi_n\left(t\right)=\phi\left(t\right)+\delta\varphi_n\left(t\right)$.
Next, we obtain the corresponding equations for small deviations $\delta\varphi_n$:
\begin{equation}
\begin{gathered}
\delta\ddot{\varphi}_n\!+\!\lambda\delta\dot{\varphi}_n\!+\!\cos\phi\!\left(t\right)\delta\varphi_n\!=\!K\left(\delta\varphi_{n\!-\!1}\!-\!2\delta\varphi_n\!+\!\delta\varphi_{n\!+\!1}\right)
\end{gathered}
\label{eq:PerturbedChainLocal}
\end{equation}
with boundary conditions $\delta\varphi_0 = \delta\varphi_1$ and $\delta\varphi_{N+1} = \delta\varphi_{N}$.
We represent the system~\eqref{eq:PerturbedChainLocal} in vector form:
\begin{equation}
\delta\ddot{\bm{\varphi}}+\lambda\delta\dot{\bm{\varphi}}+\cos\phi\left(t\right)\delta\bm{\varphi}=K\bm{A}\cdot\delta\bm{\varphi},
\label{eq:PerturbedChainLocalVector}
\end{equation}
where vector $\delta\bm{\varphi}=\left(\delta\varphi_1,\delta\varphi_2,\ldots,\delta\varphi_N\right)$, and the matrix $\bm{A}$ has the form
\begin{equation}
\bm{A}=\begin{pmatrix}
-1 &  1 & 0 &   &   &    &  0 \\
1 & -2 &  1 &   &   &    &    \\
&    &  . & . & . &    &    \\
&    &    &   & 1 & -2 &  1 \\
0 &    &    &   &   &  1 & -1 
\end{pmatrix}.
\label{eq:MatrixA}
\end{equation}
Turning on the system~\eqref{eq:PerturbedChainLocalVector} to the normal coordinates $\bm{\psi}=\left(\psi_1,\psi_2,\ldots,\psi_N\right)$, we get:
\begin{equation}
\ddot{\bm{\psi}}+\lambda\dot{\bm{\psi}}+\cos\phi\left(t\right)\bm{\psi}=K\bm{D}\cdot\bm{\psi},
\label{eq:NormalModesVector}
\end{equation}
where $\bm{D}=\text{diag}\left\{\mu_1,\mu_2,\ldots,\mu_N\right\}$ is the matrix, on the diagonal of which the eigenvalues $\mu_n$
\begin{equation}
	\mu_n=-2\left[1+\cos\left(n\pi/N\right)\right]
\end{equation}
of the matrix $\bm{A}$ are located (see Appendix~\ref{sec:Eigenproblem}). The system~\eqref{eq:PerturbedChainLocal} is divided into $N$ independent relations for normal coordinates:
\begin{equation}
\ddot{\psi}_n+\lambda\dot{\psi}_n+\left[\cos\phi\left(t\right)-K\mu_n\right]\psi_n=0.
\label{eq:NormalMode}
\end{equation}
The set of $N$ normal modes $\left\{\psi_n\right\}$ determines the stability of the in-phase mode $\phi\left(t\right)$.
The equations~\eqref{eq:NormalMode} were considered in the context of issues of stability of rotational modes in~\cite{Bolotov-19}.
It was shown that, within the limits of small dissipation $\lambda$, the trivial solution of the equation
\begin{equation}
\ddot{\psi} + \lambda \dot{\psi}+\left[\cos\phi\left(t\right)+K^{*}\right]\psi = 0,
\nonumber
\end{equation}
when the in-phase rotational motion $\phi(t)$ can be represented as an asymptotic expansion
\begin{equation}
\phi(\tau)\!=\!\tau\!+\!\dfrac{\lambda^2}{\gamma^2}\sin\tau\!+\!o\!\left(\lambda^4\right), \quad \tau \!=\! \left(\!\!\dfrac{\gamma}{\lambda} \!-\! \dfrac{\lambda^3}{2\gamma^3} \!+\! o\!\left(\lambda^7\right)\!\!\!\right)\!t,
\label{eq_rot_cycle_phase}
\end{equation}
loses its stability when $K^{*}\in\left(K_1^{*}\left(\lambda,\gamma\right), K_2^{*}\left(\lambda,\gamma\right)\right)$, and the asymptotic expressions for the values of the boundaries of the interval $K_1^{*}$ and $K_2^{*}$ have the form
\begin{equation}
K_{1,2}^{*}=\frac{1}{4}\left[\frac{\gamma^2}{\lambda^2}\mp2\sqrt{1-\gamma^2}+\frac{1}{2}\frac{\lambda^2}{\gamma^2}\right]+O\left(\frac{\lambda^4}{\gamma^4}\right).
\label{eq:K12a}
\end{equation}
Then it is easy to find
\begin{equation}
K^{\left(n\right)}_{1,2}=-K_{1,2}^{*}/\mu_n.
\label{eq:K12n}
\end{equation}
Note that one of the eigenvalues of the matrix $\bm{A}$ is equal to zero ($\mu_N=0$). The corresponding mode $\psi_N$ satisfies the equation
\begin{equation*}
\ddot{\psi}_N+\lambda\dot{\psi}_N+\cos\phi\left(t\right)\psi_N=0.
\end{equation*}
Since $\phi\left(t \right)$ is periodic, the $\psi_N$ mode is stable for any values of $K$.
Thus, there are $N-1$ intervals of values of the parameter $K$, within which $\phi\left(t \right)$ may become unstable. When the stability of the in-phase mode is lost at $K\in(K_1^{\left(n\right)}, K_2^{\left(n\right)})$, a regime arises that corresponds to the unstable mode $\psi_n$.
\par
\begin{figure*}[h!]
	% Use the relevant command to insert your figure file.
	% For example, with the graphicx package use
	\includegraphics[width=1.0\textwidth]{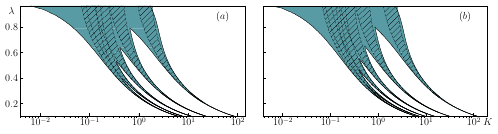}
	% figure caption is below the figure
	\caption{Regions of stability (light regions) and instability (dark regions) of the in-phase mode $\bm{\phi\left(t \right)}$ on the plane $(\lambda, K)$, determined numerically for the system~\eqref{eq:ModelGeneral} for $\gamma = 0.97$. The shaded area indicates the area of the in-phase mode instability, determined by the asymptotic boundaries of the instability regions $K_{1,2}^{(n)}$ (dashed lines), given by the expression~\eqref{eq:K12a} and \eqref{eq:K12n}. (a) $N=6$, (b) $N=7$}
	\label{fig:67pend}       % Give a unique label
\end{figure*}

Fig.~\ref{fig:67pend} shows the numerically constructed in-phase regime stability maps on the parameter plane $\left(K,\lambda\right)$ for chains at $N=6$ and $N=7$.
Theoretical analysis shows a high degree of accuracy boundary phase instability rotational mode in the case of small dissipation.
For small values $\lambda$, the number of instability regions is $N-1$. Specific rotatory structure corresponds to each such region (see the description below). It can be seen that as the $\lambda$ parameter increases, the instability regions begin to merge with each other until they turn into one.
\par
\begin{figure*}[h!]
	% Use the relevant command to insert your figure file.
	% For example, with the graphicx package use
	\includegraphics[width=1.0\textwidth]{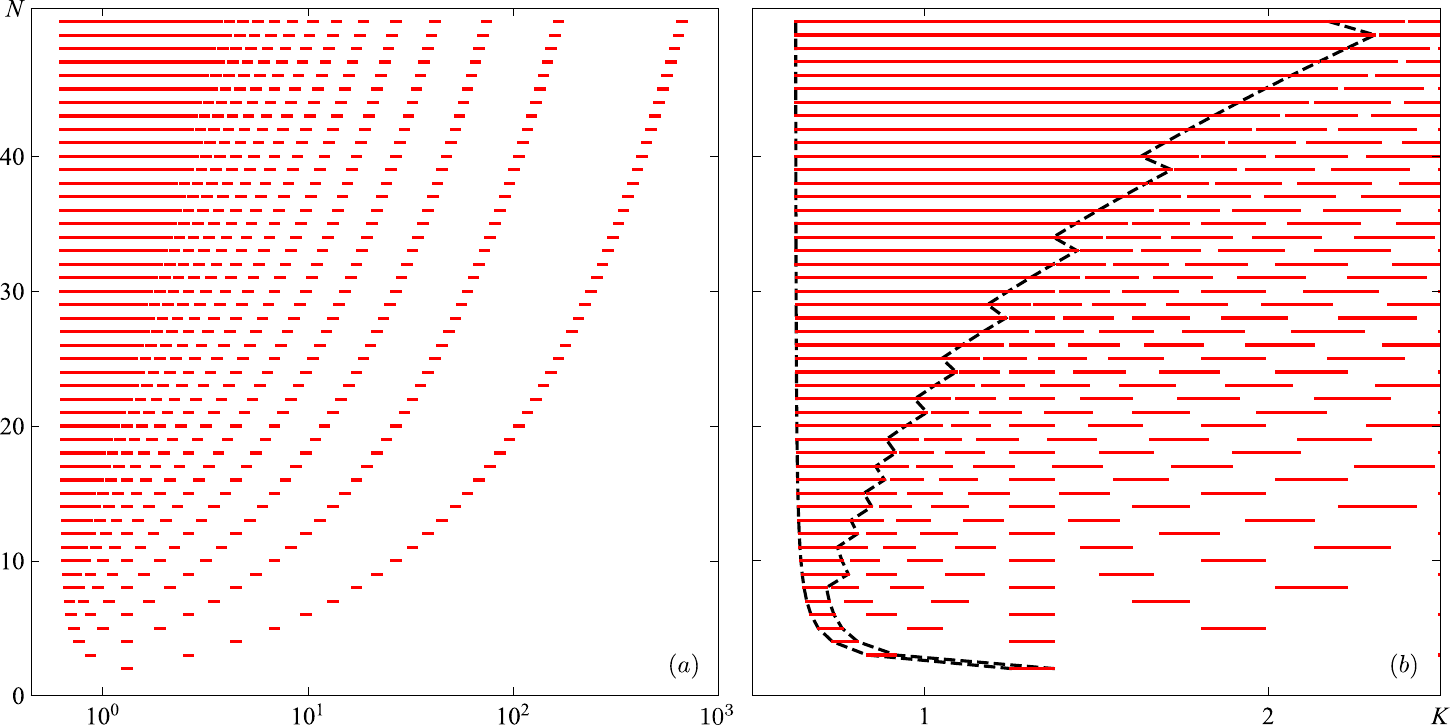}
	% figure caption is below the figure
	\caption{(a, b) Instability regions of the in-phase mode at different values of $K$. $\lambda=0.3$, $\gamma = 0.97$, $K_1^{*}\left(\lambda,\gamma\right)\approx2.5$ and $K_2^{*}\left(\lambda,\gamma\right)\approx2.75$. Blue curves show the left and right boundaries of the instability area formed by the intersection of the leftmost instability regions $\left(K_1^{\left(n\right)},K_2^{\left(n\right)}\right)$. Instability intervals are highlighted in red. Panel (b) presents an enlarged portion of the figure from the panel (a). The black dotted line indicates the right border of the instability area formed by the overlaps}
	\label{fig:UnstabilityZoom}       % Give a unique label
\end{figure*}

A similar behavior of instability regions is observed with an increase in the number of pendulums $N$. In Fig.~\ref{fig:UnstabilityZoom} on the $(K,N)$ plane, the in-phase mode instability zones are marked in red. Adding a pendulum to the system leads to the appearance of a new instability region, the zone containing the instability region $ \left(K_1^{\left(n\right)}, K_2^{\left(n\right)}\right)$ expands. The left border of this zone, defined by the expression $\min\limits_{n \in \left\{1, \ldots, N\right\}} K_1^{\left(n\right)} = K_1^{*}/\left(4\cos^2\left(\dfrac {\pi}{2N} \right) \right)$, moves to the left. The right border of this area, defined as $\max\limits_{n\in\left\{1,\ldots, N\right\}}K_2^{\left(n\right)}=K_2^{*}/\left(4\sin^2\left(\dfrac{\pi}{2N}\right)\right)$, is moved to the right. For large $N$, the value of the right boundary is equivalent to $\sim N^2$. Overlap of the instability regions located on the left begins with an increase in $N$, and the instability zone formed by the overlaps grows. This area occupies the infinite interval $\Big[L^{*}/4,+\infty\Big)$ in the limit $N \to \infty$.
Fig.~\ref{fig:UnstabilityZoom}(b) shows that with an increase in the number of elements $N$ per unit, one new zone of instability of the in-phase regime appears. Out-of-phase rotations are realized in these zones, which are structurally cluster regimes of various types. The types of rotational modes arising in the system will be denoted as $(a_1:a_2:\ldots:a_m:\ldots: a_M)$, where the number $a_m$ shows the number of elements with the same phase in the $m$-th cluster. The in-phase mode in a chain of $N$ elements is denoted by ($N:0$). Regular rotational modes are found numerically using the procedure for searching for closed trajectories (on a cylinder) of multidimensional dynamical systems (see details in~\cite{Bolotov-19}).

\section{Out-of-phase rotational modes. Cluster synchronization. Classification}
\label{sec:OutOfPhaseModes}
\begin{figure*}[h!]
	% Use the relevant command to insert your figure file.
	% For example, with the graphicx package use
	\includegraphics[width=0.65\textwidth]{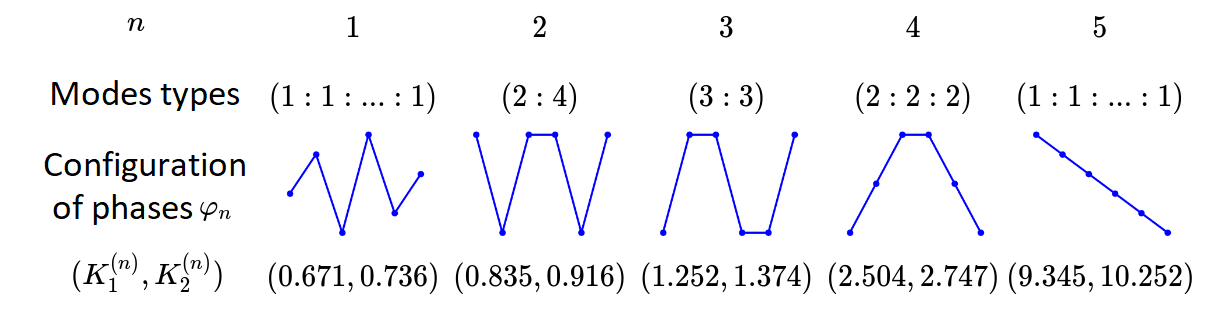}
	% figure caption is below the figure
	\caption{Configurations of out-of-phase rotational modes within regions of instability \bm{$\phi\left(t\right)}$. Parameters: $N=6$, $\lambda=0.3$, $\gamma=0.97$}
	\label{fig:6analitics}       % Give a unique label
\end{figure*}

\begin{figure*}[h!]
	% Use the relevant command to insert your figure file.
	% For example, with the graphicx package use
	\includegraphics[width=0.65\textwidth]{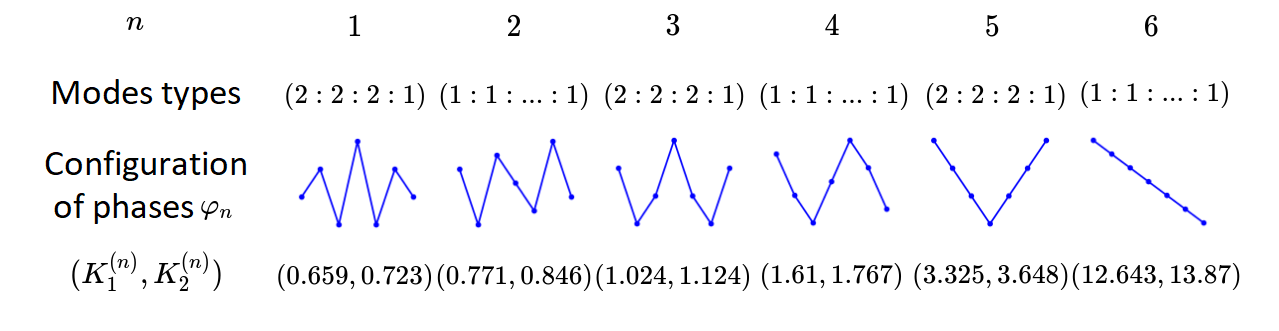}
	% figure caption is below the figure
	\caption{Configurations of out-of-phase rotational modes within regions of instability \bm{$\phi\left(t\right)}$. Parameters: $N=7$, $\lambda=0.3$, $\gamma=0.97$}
	\label{fig:7analitics}       % Give a unique label
\end{figure*}

Figs.~\ref{fig:6analitics} and \ref{fig:7analitics} show configurations of out-of-phase rotational modes for a chain of $N=6$ and $N=7$ elements, respectively. For example, the configuration $(2:2:2)$ means a phase matching mode implementation in pairs. Note, that configurations with the same $a_j$ can be different. For example, we have three different structures $(1:1:\ldots:1)$ and $(2:2:2:1)$ for $N=7$ (see Fig.~\ref{fig:7analitics}). For a given number of elements $N$, the number of synchronous clusters $M_n\left(N \right)$ of the out-of-phase mode corresponding to the $n$-th zone of instability ($n=1, 2, \dots, N-1$) is determined by the expression
	\begin{equation}
	M_n\left(N\right)=\bigg\lfloor\frac{N}{\gcd\left(2N,N-n\right)}+\frac{1}{2}\bigg\rfloor,
\end{equation}
where $\lfloor \ldots \rfloor$ is rounding down and $\gcd$ is the greatest common divisor (see Appendix~\ref{sec:Typemode}).
\par
In Fig.~\ref{fig:Theory} the boundaries of the instability regions of the in-phase regime and the types of out-of-phase rotational regimes realized in this case at fixed values of the parameters are presented based on the analysis of the expressions \eqref{eq:K12a}, \eqref{eq:K12n}.
It can be seen that both completely out-of-phase mode $(1:1:\ldots:1)$ and cluster synchronization modes (for example, $(4:4)$ for $N=8$) can be established in chains.
Analyzing the scheme in Fig.~\ref{fig:Theory}, one can see that in the case of a small dissipation at a fixed $K$, which is in the region of instability of the in-phase rotational motion, with an increase in the number of oscillators by an integer number of times, the out-of-phase rotational mode will appear again. Moreover, each of the clusters has twice as many elements.
For example, the instability region corresponding to the $(2:1)$ regime in the chain with $N=3$ again exists at $N=6$ for the $(4:2)$ regime.
\par
Below we present the results of computational experiments.
Let us introduce the synchronous parameter
\begin{equation}
\Xi=\dfrac{1}{N\left(N-1\right)}\sum\limits_{n_1,n_2=1}^{N}\smash{\displaystyle\max_{0\leq t\leq T}}\left|\dot{\varphi}_{n_1}\left(t\right)-\dot{\varphi}_{n_2}\left(t\right)\right|,
\label{eq_xi}
\end{equation}
which characterizes the degree of phase synchronization of the rotatory motion.
The value $\Xi = 0 $ shows that the rotational regime under consideration is in-phase, the values $\Xi>0 $ indicate the observation of an out-of-phase rotational regime.
For a more detailed analysis, let us present bifurcation diagrams, as well as figures showing the local maxima of the oscillator frequencies and the synchronism parameter of regular rotational modes.
\par
\begin{figure*}
	% Use the relevant command to insert your figure file.
	% For example, with the graphicx package use
	\includegraphics[width=1.0\textwidth]{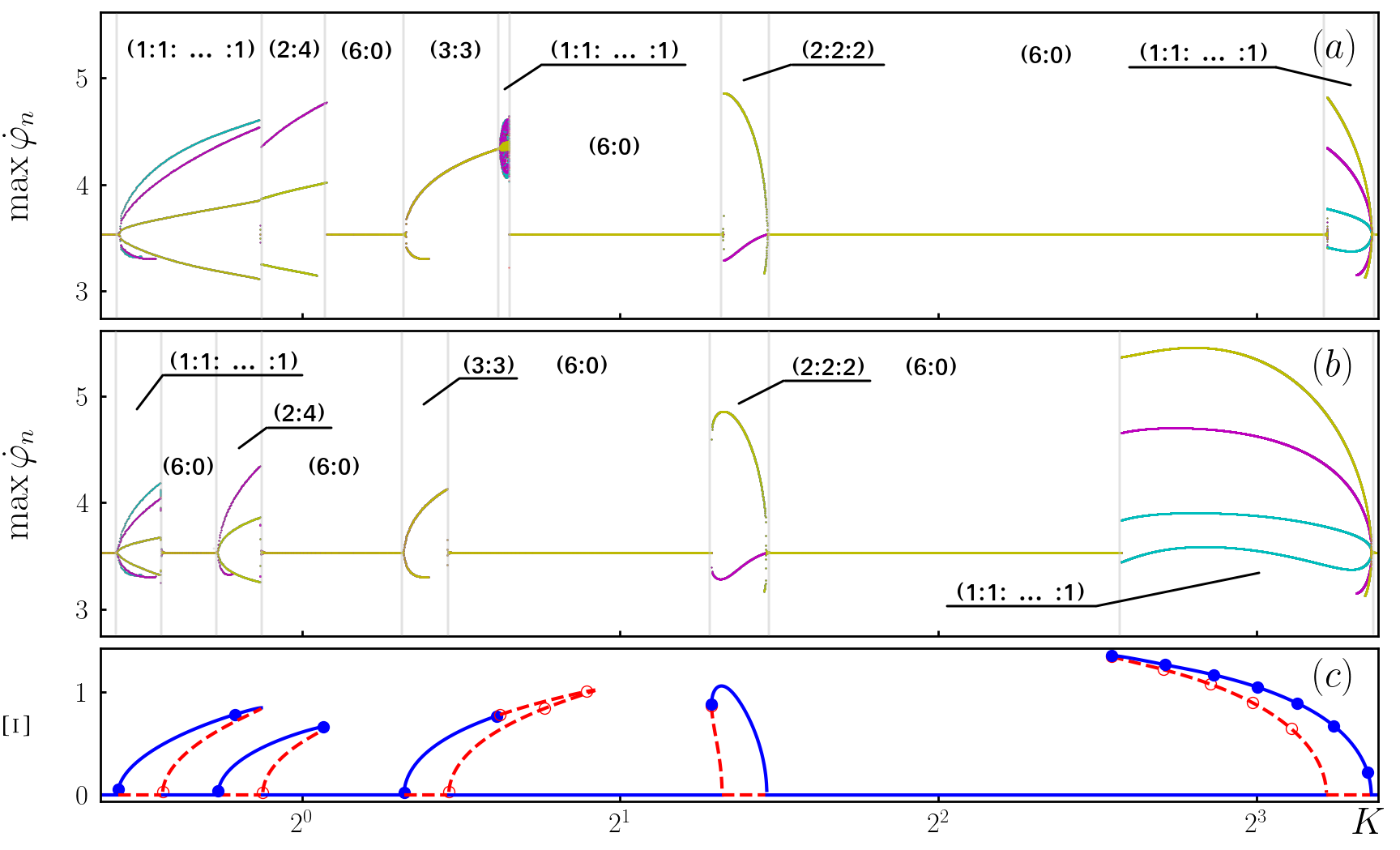}
	% figure caption is below the figure
	\caption{(a, b) Local frequency maxima. Figures are obtained by inheriting the initial conditions: (a) with increasing parameter $K$, (b) with decreasing parameter $K$. (c) Bifurcation diagram of periodic rotational regimes. $\Xi$ -- synchronous parameter, $\max{\dot{\varphi}_n}$ -- the local pendulum frequency maxima. Circular markers show $4\pi$-periodic rotational modes. Shaded markers correspond to stable rotational modes, hollow markers -- to unstable ones. The line without markers corresponds to the in-phase $2\pi$-periodic rotational regime, the solid line -- to the stable one, and the dotted line -- to the unstable one. Parameters: $N=6$, $\gamma=0.97$, $\lambda= 0.3$}
	\label{fig:bif03}       % Give a unique label
\end{figure*}

Let us analyze in details the dynamics of rotational regimes in a chain of $N=6$ pendulums depending on the parameter $K$. We consider the value of the parameter $\lambda=0.3$ (see Fig.~\ref{fig:bif03}). We are interested in the range of values of the parameter $K$, at which the rotational motion $(1: 1:\ldots:1) $ is realized for the first time. As the parameter $K$ increases, the in-phase periodic rotational motion $\phi\left(t\right) $ undergoes a period doubling bifurcation at ($K\approx 0.668$). In this case, stable in-phase $2\pi$-periodic motion gives rise to stable $4\pi$-periodic motion $(1:1:\ldots:1)$, and $2\pi$-periodic in-phase motion loses its stability. The bifurcation diagram shows (see Fig.~\ref{fig:bif03}~c) that there is also an unstable out-of-phase $4\pi$-periodic motion $(1:1:\ldots:1)$, which arises from in-phase unstable $2\pi$-periodic motion as a result of subcritical period doubling bifurcation ($K\approx 0.736$). Note that in this case the in-phase $2\pi$-periodic motion becomes stable again. Further, as the parameter $K$ increases, stable and unstable $4\pi$-periodic rotational motions $(1:1:\ldots:1)$ merge and disappear as a result of saddle-node bifurcation ($K \approx 0.911$). The same bifurcations occur with rotational motions such as $(2:4)$, $(2:2:2)$, $(1:1:\ldots:1)$. Note that a Neimark-Sacker bifurcation ($K\approx 1.543$) also appears in the system, as a result of which the $4 \pi$-periodic rotational regime $(3:3)$ loses its stability. The closed periodic trajectory turns into a torus, which becomes unstable with a further increase in the parameter $K$, and then unstable $4\pi $-periodic rotational motions merge and disappear as a result of saddle-node bifurcation ($K\approx 1.894$). We also note that for $K\in\left(0.83,0.915\right)$ in the system there is bistability of out-of-phase rotational motions, as a result of which $\left(1:1:\ldots:1\right)$ or $\left(2:4\right)$ $4\pi$-periodic rotational motions are realized depending on the initial conditions.
\par
By increasing the number of interacting elements $N$, as follows from the above analytical results, the spatiotemporal dynamics ensemble complicated. However, as our computational experiments have shown, the main bifurcations of periodic rotational regimes found for a chain of $N=6$ pendulums remain the same.

%\paragraph{Paragraph headings} Use paragraph headings as needed.
%\begin{equation}
%a^2+b^2=c^2
%\end{equation}

%% For one-column wide figures use
%\begin{figure}
%% Use the relevant command to insert your figure file.
%% For example, with the graphicx package use
%  \includegraphics{example.eps}
%% figure caption is below the figure
%\caption{Please write your figure caption here}
%\label{fig:1}       % Give a unique label
%\end{figure}
%
% For two-column wide figures use
%
% For tables use
%\begin{table}
%% table caption is above the table
%\caption{Please write your table caption here}
%\label{tab:1}       % Give a unique label
%% For LaTeX tables use
%\begin{tabular}{lll}
%\hline\noalign{\smallskip}
%first & second & third  \\
%\noalign{\smallskip}\hline\noalign{\smallskip}
%number & number & number \\
%number & number & number \\
%\noalign{\smallskip}\hline
%\end{tabular}
%\end{table}

\section{Conclusion}
\label{sec:Conclusion}
The paper considers collective rotational dynamics in a chain of locally coupled identical pendulum-type elements. It is shown that the always existing (due to the elements identity) in-phase synchronous mode in the fixed length chains loses its stability with the coupling parameter increasing at some coupling parameter intervals, which are analytically determined in the small dissipation limit. The number of such intervals is one less than the number of chain elements. In the in-phase synchronous mode instability intervals various (completely out-of-phase or cluster in-phase) frequency-synchronous modes are realized. It is shown that the evolutions hierarchy of in-phase mode into out-of-phase, then out-of-phase into in-phase mode, then (if the number of elements is $N>3$) from in-phase to out-of-phase ones, etc. with the coupling parameter increasing unambiguously depends on the chain elements number, namely, whether this number is prime, even or odd. Thus, our analysis shows that it is predictable which type of synchronous (in-phase or out-of-phase: cluster or completely out-of-phase) mode will be realized. In addition, it was shown that there are bi- and multistability regions of in-phase and out-of-phase modes. It is noteworthy that the in-phase synchronous mode stability region decreases with the elements number and dissipation increasing. A detailed system under consideration study showed that both in-phase and out-of-phase modes stability loss occurs through a period-doubling bifurcation (both direct and reverse). In the strong dissipation limit, as a result of periodic motions period-doubling bifurcations cascade, chaotic rotations arise.

%\begin{acknowledgements}
%If you'd like to thank anyone, place your comments here
%and remove the percent signs.
%\end{acknowledgements}

% Authors must disclose all relationships or interests that 
% could have direct or potential influence or impart bias on 
% the work: 
%
\section*{Conflict of interest}
The authors declare that they have no conflict of interest.

\appendix
\section{Eigenvalues and eigenvectors of the matrix $A$}
\label{sec:Eigenproblem}
To solve the eigenvalues and eigenvectors problem of the matrix \eqref{eq:MatrixA}, we perform the equivalence transformation using the upper triangular matrix $\bm{S}$ with unit elements:
\begin{equation}
\bm{S}=\left(
\begin{array}{ccccccc}
1 & 1 &   &   &   & 1 \\
0 & 1 & 1 &   &   &   \\
& 0 & . & . &   &   \\
&   & . & . & . &   \\
&   &   & . & . & 1 \\
0 &   &   &   & 0 & 1 \\
\end{array}
\right).
\label{eq:MatrixS}
\end{equation}
Then the equivalent matrix $\tilde{\bm{A}}=\bm{S}^{-1}\bm{A}\bm{S}$ equals
\begin{equation}
\tilde{\bm{A}}= \left(
\begin{array}{ccccccc}
-2&  1 &   &   &   &    & 0 \\
1& -2 & 1 &   &   &    &   \\
&  1 & . & . &   &    &   \\
&    & . & . & . &    &   \\
&    &   & . & . &  1 &   \\
&    &   &   & 1 & -2 & 0 \\
0 &    &   &   &   &  1 & 0 
\end{array}
\right),
\end{equation}
where the top left block of size $\left(N-1\right)\times\left(N-1\right)$ is tridiagonal Toeplitz matrix. Using the known results for the tridiagonal Toeplitz matrix \cite{Noschese-12} we find eigenvalues $\mu_n$ and eigenvectors $\tilde{\bm{v}}_n$ of the matrix $\tilde{\bm{A}}$:
\begin{equation}
\begin{aligned}
\mu_n&=-2\left[1+\cos\left(\frac{n\pi}{N}\right)\right],\quad n=1, 2, \dots, N-1.\\
\left[\tilde{\bm{v}}_n\right]_k&=\begin{cases} 
\left(-1\right)^{k+1}\sin\left(\dfrac{nk\pi}{N}\right),\quad \text{if}\ k=1, 2, \dots, N-1,\\ 
\dfrac{\left(-1\right)^{N-n}}{2}\tan\left(\dfrac{n\pi}{2N}\right),\quad \text{if}\ k=N, \end{cases}\\
\mu_N&=0,\quad \tilde{\bm{v}}_N=\left(0,0,\ldots,1\right).\\
\end{aligned}
\end{equation}
The eigenvalues of the origin matrix \eqref{eq:MatrixA} are also equal to $\mu_n$ and the eigenvectors $\bm{v}_n$ are defined as $\bm{v}_n=\bm{S}\tilde{\bm{v}}_n$:
\begin{equation}
\begin{aligned}
\left[\bm{v}_n\right]_k&\!=\!\left(\!-1\!\right)^{k\!+\!1}\frac{\sin\!\left[\left(2k\!-\!1\right)\!\dfrac{n\pi}{2N}\right]}{2\cos\!\left(\dfrac{n\pi}{2N}\right)},\quad \!\!\!\!\!\!n=1,\dots, N\!-\!1,\quad \!\!\!\!\!\!k=1,\dots, N,\\
\bm{v}_N&=\left(1,1,\ldots,1\right).
\end{aligned}
\end{equation}
Performing the eigenvectors $\bm{v}_n$ normalization for $n=1,\dots,N-1$, we write the final result in general form:
\begin{align}
\mu_n&=-2\left[1+\cos\left(\frac{n\pi}{N}\right)\right],&n=1,\dots,N,\label{eq:mu_n}\\
\left[\bm{v}_n\right]_k&=\left(-1\right)^{k+1}\sqrt{\frac{2}{N}}\sin\left[\left(2k-1\right)\frac{n\pi}{2N}\right],&k={1,\dots,N}.\label{eq:v_n_k}
\end{align}
(with this form, the vector $\bm{v}_N$ remains unnormalized).

\section{Determing the type of out-of-phase mode}
\label{sec:Typemode}
There is a one-to-one relationship between the eigenvectors $\bm{v}_n$ form and the emerging $4\pi$-periodic out-of-phase rotations, which makes it possible to determine the number and size of synchronous clusters. Thus, the number of clusters $M_n\left(N\right)$ of the out-of-phase mode arising as a result of the $\psi_n$ ($n={1, \dots, N-1}$) mode instability development is equal to the number of different values that the components of the vector $\bm{v}_n$ take, i.e.
\begin{equation}
M_n\left(N\right)=\left|\left\{\left[ v_n\right]_1,\ldots,\left[v_n\right]_N\right\}\right|.
\end{equation}
In addition, the number of elements $N_m^{\left(n\right)}\left(N\right)$ in the $m$-th cluster ($m=1, \dots, M_n$) is equal to the number of the vector $\bm{v}_n$ components taking some value from the set $\left\{\left[ v_n\right]_1,\ldots,\left[v_n\right]_N\right\}$. To determine the numbers $M_n$ and $N_m^{\left(n\right)}$, consider a sequence whose elements are specified by the expression \eqref{eq:v_n_k} without restriction on the index $k$: ${\ldots,\left[v_n\right]_{-1},\left[v_n\right]_0,}$
$\left[v_n\right]_1,{\left[v_n\right]_2,\ldots}$. Rewriting \eqref{eq:v_n_k} as
\begin{equation}
\left[\bm{v}_n\right]_k=\sqrt{\frac{2}{N}}\cos\left[\left(2k-1\right)\frac{\left(N-n\right)\pi}{2N}\right],
\end{equation}
from the cosine function periodicity, it is easy to find that the introduced sequence has a period $\tilde{k}=2N/\gcd\left(2N,N-n\right)$: $\left[v_n\right]_{k+\tilde{k}}=\left[v_n\right]_k$. It is easy to see that $3\le\tilde{k}\le 2N$. In addition, the symmetry of the cosine function implies that $\left[v_n\right]_0=\left[v_n\right]_1$. These two properties of the sequence $\left\{\left[v_n\right]_k\right\}_{k=-\infty}^{+\infty}$ elements allow them to be divided into groups with equal values. From the periodicity property it follows that all elements with unique values are contained in a subsequence $\left[v_n\right]_1,$ $\left[v_n\right]_2, \ldots, \left[v_n\right]_{\tilde{k}}$. The symmetry condition entails that $\left[v_n\right]_1=\left[v_n\right]_{\tilde{k}},$ $\left[v_n\right]_2=\left[v_n\right]_{\tilde{k}-1}$, etc.
Thus, the number of unique in value elements, that is, the number of synchronous clusters, is $M_n\left(N\right)\!=\!\Big\lfloor\!\!\left(\tilde{k}+1\right)/2\!\Big\rfloor$, or explicitly
\begin{equation}
M_n\left(N\right)=\bigg\lfloor\frac{N}{\gcd\left(2N,N-n\right)}+\frac{1}{2}\bigg\rfloor.
\end{equation}
From the period value $\tilde{k}$ constraints it follows, that $2\le M_n\left(N\right)\le N$, moreover $M_n\left(N\right)=N$ if and only if $2N$ and $N-n$ are coprime integers.

Further, the value $\left[v_n\right]_m$ ($m=1, \dots, M_n$) for $k=1, \dots, N$ occurs $\left(\Big\lfloor\left(N-m\right)/\tilde{k}\Big\rfloor+1\right)+\left(\Big\lfloor\left(N-\left(\tilde{k}-m+1\right)\right)/\tilde{k}\Big\rfloor+1\right)$ times if condition $m\ne\left(\tilde{k}+1\right)/2$ satisfied, and $\left(\Big\lfloor\left(N-m\right)/\tilde{k}\Big\rfloor+1\right)$ times when $m=\left(\tilde{k}+1\right)/2$. After elementary simplifications, we have
\begin{equation}
N_m^{\left(n\right)}\left(N\right)=\frac{1}{1+\delta_{2m,\tilde{k}+1}}\left(\bigg\lceil\frac{N-m+1}{\tilde{k}}\bigg\rceil+\bigg\lfloor\frac{N+m-1}{\tilde{k}}\bigg\rfloor\right),
\end{equation}
$ m={1, \dots, M_n\left(N\right)}$.
Substituting the period $\tilde{k}$ and making simplifications, we finally find
\begin{equation}
N_m^{\left(n\right)}=\begin{cases} \gcd\left(2N,N-n\right), & \mbox{if } m\ne\frac{N}{\gcd\left(2N,N-n\right)}+\frac{1}{2}, \\ \gcd\left(2N,N-n\right)/2, & \mbox{else.} \end{cases}
\end{equation}

From the latter expression follows a simple rule describing the cluster regime arising in connection with the $\psi_n$ mode instability. This regime has $M_n\left(N\right)$ clusters (see above). If the condition $\dfrac{N}{\gcd\left(2N,N-n\right)}+\dfrac{1}{2}\notin\mathbb{Z}$ is satisfied (or equivalently, $N/\gcd\left(2N,N-n\right)=M_n\left(N\right)$), then all $M_n\left(N\right)$ clusters contain $\gcd\left(2N,N-n\right)$ elements. Otherwise, $M_n\left(N\right)\!-\!1$ clusters contain $\gcd\left(2N,N-n\right)$ elements each and one cluster contains $\gcd\left(2N,N-n\right)/2$ elements.

% BibTeX users please use one of
%\bibliographystyle{spbasic}      % basic style, author-year citations
%\bibliographystyle{spmpsci}      % mathematics and physical sciences
%\bibliographystyle{spphys}       % APS-like style for physics
%\bibliography{}   % name your BibTeX data base

% Non-BibTeX users please use
%\begin{thebibliography}{}
%%
%% and use \bibitem to create references. Consult the Instructions
%% for authors for reference list style.
%%
%\bibitem{RefJ}
%% Format for Journal Reference
%Author, Article title, Journal, Volume, page numbers (year)
%% Format for books
%\bibitem{RefB}
%Author, Book title, page numbers. Publisher, place (year)
%% etc
%\end{thebibliography}

\begin{figure*}[h!]
	% Use the relevant command to insert your figure file.
	% For example, with the graphicx package use
	\includegraphics[width=1.0\textwidth]{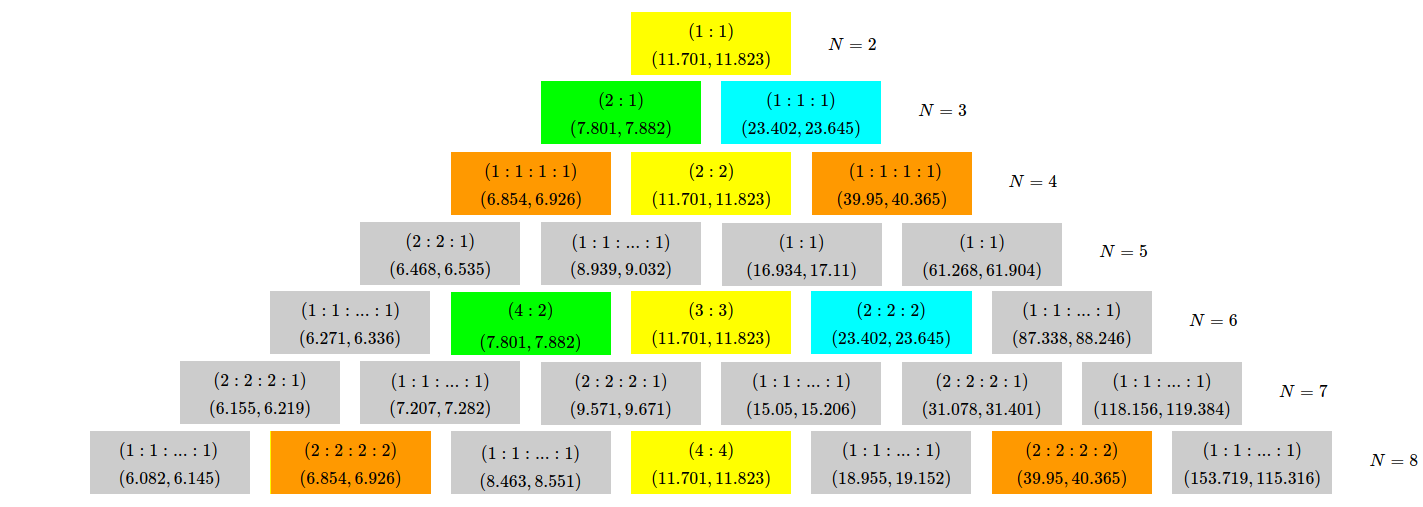}
	% figure caption is below the figure
	\caption{Stable out-of-phase rotational modes, in chains with different numbers of pendulums $N$. Each cell shows the range of values of the coupling parameter $ K $, within which the in-phase rotational mode is unstable \eqref{eq:K12n} and also the type of out-of-phase rotational mode. Cells of the same color, except for gray ones, indicate the same range of the instability region in terms of the $K$ parameter. Parameters: $\gamma=0.97$, $\lambda=0.1$}
	\label{fig:Theory}       % Give a unique label
\end{figure*}

\end{document}